\documentclass[Journal,letterpaper]{ascelike-new}
\usepackage[utf8]{inputenc}
\usepackage[T1]{fontenc}
\usepackage{lmodern}
\usepackage{graphicx}
\usepackage[figurename=Fig.,labelfont=bf,labelsep=period]{caption}
\usepackage{amsmath}
\usepackage{subfigure}
\usepackage{newtxtext,newtxmath}
\usepackage[colorlinks=true,citecolor=red,linkcolor=black]{hyperref}
\usepackage[flushleft]{threeparttable}
\usepackage{multirow}
\usepackage[ruled,vlined]{algorithm2e}
\NameTag{Wang et al. \today}
\begin{document}

% You will need to make the title all-caps
\title{Minimizing Pumping Energy Cost in Real-time Operations of Water Distribution Systems using Economic Model Predictive Control}

\author{Ye Wang\thanks{Research Fellow, PhD, Department of Electrical and Electronic Engineering, The University of Melbourne, VIC 3010, Australia}, Kevin Too Yok\thanks{PhD Student, B.Eng., Department of Electrical and Electronic Engineering, The University of Melbourne, VIC 3010, Australia}, Wenyan Wu\thanks{Senior Lecturer, PhD, Department of Infrastructure Engineering, The University of Melbourne, VIC 3010, Australia and Adjunct Lecturer, School of Civil, Environmental \& Mining Engineering, University of Adelaide, SA 5005, Australia}, Angus R. Simpson\thanks{Professor, PhD, School of Civil, Environmental \& Mining Engineering, University of Adelaide, SA 5005, Australia and Honorary Professorial Fellow, Department of Infrastructure Engineering, The University of Melbourne, VIC 3010, Australia}, Erik Weyer\thanks{Professor, PhD, Department of Electrical and Electronic Engineering, The University of Melbourne, VIC 3010, Australia}, Chris Manzie\thanks{Professor, PhD, Department of Electrical and Electronic Engineering, The University of Melbourne, VIC 3010, Australia}
}
% \affil[2]{Department of Infrastructure Engineering, The University of Melbourne, VIC 3010, Australia}
% \affil[3]{School of Civil, Environmental \& Mining Engineering, University of Adelaide, SA 5005, Australia}

\maketitle
% Please include an abstract:
\begin{abstract}
	Optimizing pump operations is a challenging task for real-time management of water distribution systems (WDSs). With suitable pump scheduling, pumping costs can be significantly reduced. In this research, a novel economic model predictive control (EMPC) framework for real-time management of WDSs is proposed. Optimal pump operations are selected based on predicted system behavior over a receding time horizon with the aim to minimize the total pumping energy cost. Time-varying electricity tariffs are considered while all the required water demands are satisfied. The novelty of this framework is to choose the number of pumps to operate in each pump station as decision variables in order to optimize the total pumping energy costs. By using integer programming, the proposed EMPC is applied to a benchmark case study, the Richmond Pruned network. The simulation with an EPANET hydraulic simulator is implemented. Moreover, a comparison of the results obtained using the proposed EMPC with those obtained using trigger-level control demonstrates significant economic benefits of the proposed EMPC.
\end{abstract}

\section{Introduction}

Water distribution systems (WDSs) are critical infrastructure for modern cities. With appropriate operational management, water companies are able to provide water with the desired quantity and quality to all the customers at a reasonable operating cost. Pumping costs constitute a significant proportion of the operating cost for WDSs, and {they are projected to increase in the future due to an increase in population leading to larger water consumption~\cite{outlookreport2016}. An increasingly important objective in real-time optimal pump operations of WDSs is therefore to minimize pumping energy costs.} During the past three decades, a large amount of research has been carried out on optimal operations of WDSs. A comprehensive literature review including over 100 scientific publications from the 1970s until 2017 on the optimization of WDSs was conducted by~\citeN{Mala-Jetmarov2017}. The 2017 study revealed that pump operation optimization is one of the two major areas of optimization research related to WDSs (the other area being water quality optimization). 

WDS operation optimization is a very difficult problem due to a number of challenges, which include but are not limited to: 1) the equations governing the hydraulic interactions within the network are nonlinear; 2) decision variables can be either continuous or discrete; and 3) real WDSs typically consist of a large number of branches, links, loops and nodes. In addition, there are new challenges for the operation of modern WDSs, such as, 4) electricity prices can change with time; 5) disturbances to the system may occur including degradation of the pipe and pump characteristics; and {6) the need to enforce constraints on water volumes and pumping flow rates within a WDS.} Consequently, to solve a pump operation problem, a mathematical model of the system is needed taking into account the nonlinear hydraulic equations and the binary (on/off) nature of pumping decisions. The resulting mixed-integer, dynamic and nonlinear optimization problem is numerically challenging to solve in real-time. As a result, researchers often make use of a number of simplifying assumptions to reduce the computational requirements, which has resulted in the development of a variety of problem formulations and solution methods over time. 

Early studies on WDS operation optimization focused on deterministic approaches, such as dynamic programming (DP) \cite{sterling_dynamic_1975}, linear programming (LP) (\citeNP{giacomello_fast_2013}; \citeNP{kurian_optimal_2018}), nonlinear programming (NLP) (\citeNP{brion_lehar_m._methodology_1991}; \citeNP{ormsbee_lindell_e._nonlinear_1995}), and mixed integer non-linear programming (MINLP) (\citeNP{biscos_optimal_2003}; \citeNP{bagirov_algorithm_2013}). Recently, meta-heuristic
algorithms, such as genetic algorithms (\citeNP{paschke2001genetic}; \citeNP{kazantzis2002new}; \citeNP{Van2004}; \citeNP{wang_enhanced_2009}; \citeNP{Wu-2010}; \citeNP{Wu-2012b}; \citeNP{Wu-2012a}; \citeNP{Lisa-Angus-2016}), simulated annealing (\citeNP{cunha_maria_da_conceicao_water_1999}; \citeNP{goldman_application_1999}), particle swarm optimization \cite{wegley_chad_determining_2000}, and ant colony optimization (\citeNP{ostfeld_avi_ant_2008}; \citeNP{lopez-ibanez_manuel_ant_2008}) have been used for finding open-loop schedules for pump operations. However, the majority of these studies did not consider real-time pump operations, where a feedback control loop uses measurements of the current system state to make informed decisions in real-time about how to choose system inputs in the future.

A well-known advantage of closed-loop controllers is robustness to system modeling errors although they require real-time measurements of system states and increased online computational resources. Whilst a wide range of feedback control algorithms are in use, model predictive control (MPC) has gained increasing popularity over the past decades due to its ability to explicitly handle system constraints and deliver improved system performance through the use of a model of the system.  It is also readily deployable to multi-input-multi-output systems.

MPC was initially developed for the chemical processing industry in the late 1970s~\cite{Garcia-1989}. It relies on an internal model of the process dynamics to predict system states and outputs given a sequence of control actions over a finite receding horizon. An optimization problem is solved at each sampling instant to determine an optimal control sequence that minimizes deviations of the process states from some predetermined operating setpoints. MPC utilizing a quadratic cost on state deviations and control actions has been applied to many water applications, such as river management and operation of open channel systems~(\citeNP{Nasir-Erik-2018}; \citeNP{Nasir-Erik-2020}; \citeNP{Nasir-Erik-2019}). 

Economic MPC (EMPC) is a relatively new extension of MPC, where the cost function is generally formulated using an arbitrary objective function that captures some economic aspect of process systems to be controlled (\citeNP{rawlings_fundamentals_2012}; \citeNP{ellis_economic_2016}). Given its superior "economic" performance, EMPC has found recent applications in various areas, including building energy systems \cite{ma_demand_2012}, gas pipeline networks \cite{gopalakrishnan_economic_2013}, electric vehicle charge planning \cite{halvgaard_electric_2012}, inventory management \cite{subramanian_economic_2014}, and wastewater treatment processes \cite{zeng_economic_2015}. EMPC has been found to be a suitable control strategy for operational management of WDSs. An earlier deployment of EMPC for the water distribution problem was proposed by~\citeN{Cembrano-Wells-2000}, where the simulations only considered a two pump system model and locally optimal solutions were found. More recently, this approach has been revisited using flow-based and pressured-based models of WDSs \cite{Wang-Puig-2017}. The works by ~\citeN{Wang-Puig-2017} and ~\citeN{Salomons-Housh-2020} have suggested the use of EMPC for WDSs but the authors have made simplifications including approximating the on/off behavior of pumps, as well as using simplified plant and economic/energy models. 

This study proposes a novel EMPC framework for real-time management of WDSs, where the full nonlinear pumping models and accurate energy pricing are taken into account to investigate the benefits of EMPC relative to trigger-level control. Within this EMPC framework, a modeling methodology for WDSs is introduced, which includes the number of pumps in each pump station as integer decision variables. The advantage of this framework is to make sure that accurate economic pumping costs are included in the control system. Based on the objectives of the operational management of WDSs, economic cost functions and constraints have been formulated as part of the EMPC optimization problem. By using integer programming to solve the corresponding EMPC optimization problem online with updated feedback information, optimal pump operations can be obtained while minimizing the pumping costs. A case study for the Richmond Pruned network is presented here based on the original Richmond network introduced in the study by~\citeN{Van2004}. The closed-loop online simulation results with EMPC connected to the Richmond Pruned network implemented in the EPANET hydraulic simulator~\cite{epanet}. The simulation results also reveal a natural pump operation in view of the electricity prices and the water volumes in storage tanks. A comparison of the results with trigger-level control demonstrate the effectiveness of EMPC. 

%%%%%%%%%%%%%%%%%%%%%%%%%%%%%%%%%%%%%%%%%%%%%%%%%%%%%%%%%%%%%%%
The remainder of this paper is organized as follows. First, the EMPC framework for real-time pump operations is proposed. Second, the Richmond Pruned case study is described in detail. Then, the closed-loop online simulation results are discussed for six different demand loading cases and a comparison of the results with those obtained by trigger-level control is presented. Finally, conclusions are drawn and future research directions are suggested. 

%%%%%%%%%%%%%%%%%%%%%%%%%%%%%%%%%%%%%%%%%%%%%%%%%%%%%%%%%%%%%%%
%%%%%%%%%%%%%%%%%%%%%%%%%%%%%%%%%%%%%%%%%%%%%%%%%%%%%%%%%%%%%%%
\section{Economic Model Predictive Control for Optimal Pump Operations}\label{section:EMPC}

In this section, a novel EMPC framework is systematically described for real-time operational management of WDSs. A mathematical modeling methodology is first introduced to find the flow-based model of a WDS. Then, the cost functions and constraints are formulated. The EMPC optimization problem is then presented within this given framework. 

\subsection{WDS Model}

To design an EMPC controller, a lower-order prediction model of a WDS is required to describe the system dynamics. In this research, a flow-based model of WDSs is considered, which consists of $ T $ water storage tanks, $ P_s $ pump stations with a total of $ \overline{n}_j $ parallel pumps located in the $ j $-th pump station, and $ D $ water demands. The corresponding variable assignments are shown in Table~\ref{table:variable assignment}. 

\begin{table}
	\caption{Variable assignment for modeling.}
	\label{table:variable assignment}
	\centering
	\small
	\renewcommand{\arraystretch}{1.25}
	\begin{tabular}{l c c c}
	\hline\hline
	Item & Variable & Minimum Value & Maximum Value\\
	\hline
	Tank water depth & $ x_i $ & 0 & $ \overline{x}_i $ \\
	Inflow to the tank & $ q_i^{(in)} $ & 0 & - \\
	Outflow from the tank & $ q_i^{(out)} $ & 0 & - \\
	Number of pumps operating & $ n_j $ & 0 & $ \overline{n}_j $ \\
	Outflow from pump station & $ q_j $ & 0 & - \\
	Water demand & $d_s$ & 0 & -\\
	Subscript indexing the tanks & $ i $  & 1 & $T$ \\
	Subscript indexing the pump stations & $ j $ & 1 & $ P $\\
	Subscript indexing the water demands & $ s $ & 1 & $ D $\\
	%\hline
	%\multicolumn{2}{l}{$\ast$ $D_{50}$ represents the median particle diameter} \\
	\hline\hline
	\end{tabular}
	\normalsize
\end{table}

For each water storage tank $ i = 1,\ldots,T $, the variable $ x_i $ represents the water depth in the tank. {The water depth in a storage tank as a function of a water volume balance can be written as}
\begin{equation}\label{eq:x function}
	x_i(k+1) = x_i(k) + \frac{\Delta t}{S_i} \left( q_{i}^{(in)}(k) - q_{i}^{(out)}(k) \right) , \qquad i=1,\dots,T,\; k = 0,1,\dots, N,
\end{equation}
where $ q_{i}^{(in)} $ and $ q_{i}^{(out)} $ represent the inflow and outflow for the $ i $-th water tank, $ k $ refers to the $ k $-th time step from a total of $ N $ time steps, $ \Delta t $ is the sampling time and $ S_i $ is the area (plan view) of the $ i $-th tank. The inflow to a tank is usually determined by the outflows from pump stations while the outflow from a tank is usually determined by the water demands $ d_s $, $ s=1,\ldots, D $ from each water demand node.

For each pump station $ j = 1,\dots,P_s $ , the variable $ n_j $ represents the number of parallel pumps that are currently operating, and the variable $ q_j $ represents the total outflow from the $ j $-th pump station. {The pump flow rate is dependent on downstream hydraulic conditions such as water demand loading and tank water depths. However, since the variations in flow rates due to these factors are relatively small, approximations of the flow rates that only depends on the number of pumps operating is used, i.e.}
\begin{equation}\label{eq:q-u function}
	q_j(k) = \phi_j(n_j(k)), \qquad j = 1, \dots, P_s, \; k = 0,1,\dots, N,
\end{equation}
{where $ \phi_j(n_j) $ can be obtained from interpolated experimental or simulated (e.g. EPANET) data, noting that $ n_j $ can take integer values from 0 to $ \overline{n}_j $.}

{In addition, for nodes without storage in a water system, the sum of inflows should be equal to the sum of outflows. This can be written as
\begin{equation}
    \mathbf{0} = g (\mathbf{u}(k),\mathbf{d}(k)),\qquad k = 0,1,\dots, N,
\end{equation}
where $g (\mathbf{u}(k),\mathbf{d}(k))$ is a vector function. Each element of this vector function corresponds to a node.}

In general, the flow-based prediction model for a WDS can be summarized as follows:
\begin{subequations}\label{eq:general model}
	\begin{align}
		\mathbf{x}(k+1) &= f (\mathbf{x}(k),\mathbf{u}(k),\mathbf{d}(k)), \qquad k = 0,1,\dots, N,\label{eq:dynamical equation}\\
		\mathbf{0} &= g (\mathbf{u}(k),\mathbf{d}(k)),\qquad \qquad \; k = 0,1,\dots, N,\label{eq:algebraic equation}
	\end{align}
\end{subequations}
where $ \mathbf{x}=\left[ x_1,\ldots,x_T\right]^{\top} $, $ \mathbf{u} = \left[ n_1,\ldots,n_{P_s} \right]^{\top} $, $ \mathbf{d}=\left[ d_1,\ldots,d_D \right]^{\top} $ denote the vectors of tank water depths, the number of pumps operating in each pump station, and the water demands across the whole water network. Note that for complex WDS, Eq. \eqref{eq:dynamical equation} obtained from Eqs.~\eqref{eq:x function} and~\eqref{eq:q-u function}, and {the equation in Eq.~\eqref{eq:algebraic equation} obtained from the mass balance relationships for the nodes without storage.} In this research, the water demands are assumed to be known along a prediction horizon as considered in EMPC by using demand forecasting methods, such as the ones introduced in the studies by~\citeN{Ye-Chapter-2016} and~\citeN{Salomons-Housh-2020}.

\subsection{Cost Functions and Constraints Setup}

In the following, cost functions and constraints are defined based on the objectives for real-time operational management of WDSs.

% \subsubsection{Cost Functions}
{The power consumption~$ P $ for an operating pump is modeled as
\begin{equation}\label{eq:power}
	P = \frac{\gamma Q H_p}{\eta_s \eta_m},
\end{equation}
where $ \gamma $ is specific weight of water [N/m$^3$], $ Q $ is a pump flow [m$^3$/s], $ H_p $ is a pump head [m], $ \eta_s $ is the pump shaft efficiency and $ \eta_m $ is the motor efficiency. From Eq.~\eqref{eq:power}, the energy consumed by a pump operating for a time interval of $ \Delta t  $ is
\begin{equation}\label{eq:energy}
	E = P \Delta t.
\end{equation}}
{From Eqs.~\eqref{eq:power} and~\eqref{eq:energy}, the pump head $ H_p $ is used to compute the power and energy for a single pump, however, the flow-based model~\eqref{eq:general model} assumes pump energy usage is independent of head (which is a simplification as it ignores any relationship to the state $\mathbf{x}$). The energy consumed in the $j$-th pump station is therefore approximated as $E_j = {\psi_j}(n_j) $. Taking into account the time-varying electricity prices from the tariff database, the pumping energy cost can be computed at each sampling interval for the different prices.} For the $ j $-th pump station, given a time-varying electricity price at time step $ k $ as $ \alpha_j(k) $, the total pumping energy cost can be expressed as
\begin{equation}\label{eq:general single economic cost function}
	\ell_{j} \left( n_j(k), \alpha_j(k) \right) = \alpha_j (k) {\psi_j} \left( n_j(k) \right), \qquad j = 1, \dots, P_s, \; k = 0,1,\dots, N,
\end{equation}
where $ {\psi_j}(n_j) $ is a function that estimates the energy consumed by $ n_j $ parallel pumps at the $ j $-th pump station (assuming that all the pumps in a pump station are of the same type). 

The total economic cost function for the $ P_s $ pump stations as a function of the vector $ \mathbf{u}(k) $ of the actual number of pumps operating in each pump station can be stated as
\begin{equation}\label{eq:economic cost function}
	\ell_e(\mathbf{u}(k), \alpha (k)) = \sum_{j=1}^{P_s} \ell_{j} \left(n_j(k),\alpha_j (k) \right), \qquad k = 0,1,\dots, N,
\end{equation}
where the subscript $ e $ refers to the "economic" cost function associated with pumps that are operating and $ \alpha(k) = [\alpha_1(k),\ldots, \alpha_{P_s}(k)]^{\top} $. 

While it cannot be directly attributed to the short term operating costs of the network, excessive pump switching is considered undesirable as it can lead to mechanical degradation and early replacement (with a consequent economic impact on the long term operation of the network). As a result, an implicit economic cost that is associated with turning pumps on and off is introduced using the following economic penalty
\begin{equation}\label{eq:smoonthness cost function}
	\ell_p(\mathbf{u}(k)) = \left\| \Delta \mathbf{u}(k) \right\|_{R}^2=\Delta \mathbf{u}(k)^{\top} R \Delta \mathbf{u}(k),  \qquad k = 0,1,\dots, N,
\end{equation}
where $ \Delta \mathbf{u}(k) = \mathbf{u}(k) - \mathbf{u}(k-1) $, and $ \left\| \cdot \right\|_R  $ refers to the weighted 2-norm by a positive-definite matrix~$ R $. The weighting matrix $ R $ may be tuned to account for different types pumps in different pump stations, or the different network architecture implicitly allowing more switching behavior in some pump stations.

% \subsubsection{Constraints}
In a WDS, physical limitations on water depths in the tanks and the availability of pumps in a pump station must be taken into account. For the $ T $ water tanks in the network, the constraint describing the physical limitations or operational limits for water depths can be formulated by
\begin{equation}\label{eq:constraint-x}
	\underline{\mathbf{x}} \leq \mathbf{x}(k) \leq \overline{\mathbf{x}}, \qquad k = 0,1,\dots, N,
\end{equation}
where $ \underline{\mathbf{x}} $ and $ \overline{\mathbf{x}} $ denote vectors of lower and upper bounds for tank water depths, respectively.

For the $ P_s $ pump stations in the network, the constraints on the number of available pumps can be described by
\begin{equation}\label{eq:constraint-u}
	\mathbf{0} \leq \mathbf{u}(k) \leq \overline{\mathbf{u}}, \qquad k = 0,1,\dots, N,
\end{equation}
where $ \overline{\mathbf{u}}  = \left[ \overline{n}_1,\ldots,\overline{n}_{P_s} \right]^{\top}  $ denotes the vector of the available number of parallel pumps in each pump station. Note that the elements of the vector $ \overline{\mathbf{u}} $ take on integer values.

\subsection{Optimization Problem Formulation}

According to the discussion above, the EMPC controller for the operational management of WDSs can be implemented by solving a finite-horizon optimization problem considering {a prediction horizon $ N > 0 $}, that is, at each time $ k \geq 0 $, solve

\begin{subequations}\label{problem:EMPC}
	\begin{align}
		&\underset{\substack{\tilde{\mathbf{u}}(k),\ldots,\tilde{\mathbf{u}}(k+N-1)}} {\mathrm{minimize}}\;\; \sum_{t=k}^{k+N-1}\Big( \ell_e (\tilde{\mathbf{u}}(t)) + \ell_p(\tilde{\mathbf{u}}(t)) \Big) ,\label{eq:MPC cost function}\allowdisplaybreaks\\
		%%%%%%%%%%%%%%%%
		\text{subject } \text{to} &\nonumber \\
		& \tilde{\mathbf{x}}(t+1) = f (\tilde{\mathbf{x}}(t),\tilde{\mathbf{u}}(t),\tilde{\mathbf{d}}(t)), \qquad &&t = k,\ldots,k+N-1,\label{eq:MPC prediction model-1}\allowdisplaybreaks\\
		& \mathbf{0} = g (\tilde{\mathbf{u}}(t),\tilde{\mathbf{d}}(t)), \qquad &&t = k,\ldots,k+N-1\label{eq:MPC prediction model-2}\allowdisplaybreaks\\
		& \underline{\mathbf{x}} \leq \tilde{\mathbf{x}}(t+1) \leq \overline{\mathbf{x}}, \qquad && t = k,\ldots,k+N-1,\label{eq:MPC constraint-x}\allowdisplaybreaks\\
		& \mathbf{0} \leq \tilde{\mathbf{u}}(t) \leq \overline{\mathbf{u}}, \qquad && t = k,\ldots,k+N-1,\label{eq:MPC constraint-u}\allowdisplaybreaks\\
		& \tilde{\mathbf{x}}(k) = \mathbf{x}(k), \label{eq:MPC initialisation}\\
		& \left[ \tilde{\mathbf{d}}(k),\ldots, \tilde{\mathbf{d}}(k+N -1 )\right]^{\top} = \left[ \mathbf{d}(k),\ldots, \mathbf{d}(k+N - 1 )\right]^{\top},\label{eq:MPC demand forecast}
	\end{align}
\end{subequations}
where the tilde $ \sim $ refers to the predicted variables. Eq.~\eqref{eq:MPC cost function} is the total cost function, Eqs.~\eqref{eq:MPC prediction model-1} and \eqref{eq:MPC prediction model-2} are the constraints that the future states must obey the plant dynamics of a WDS based on the volume balance model in Eq.~\eqref{eq:general model}, Eq.~\eqref{eq:MPC constraint-x} constrains the water depth in each tank, Eq.~\eqref{eq:MPC constraint-u} constrains the number of parallel pumps that can be operated in each pump station, Eq.~\eqref{eq:MPC initialisation} is the initialization constraint to feed back the current measured water depth $ \mathbf{x}(k) $ as the first predicted state $ \tilde{\mathbf{x}}(k) $, and Eq.~\eqref{eq:MPC demand forecast} are the demand forecasts over {a prediction horizon $ N $} under the assumption of no uncertainty in the demand forecasts. 

From the solutions to the optimization problem in Eq.~\eqref{problem:EMPC} at any time $ k $, the optimal control action at sampling time $ k $ is chosen as
\begin{equation}\label{eq:optimal control action}
	\mathbf{u}(k) = \tilde{\mathbf{u}}^{*}(k),
\end{equation}
where $ \mathbf{u}^{*}(k) $ is the first value of the optimal control sequence as determined from solving Eq.~\eqref{problem:EMPC}. Thus, at each time step $ k $, a new optimization in Eq.~\eqref{problem:EMPC} is carried out for a {prediction horizon $ N $} but only the first control input is implemented. At the next time step $ k+1 $, all the system states are updated using the measured data from the WDS, and the optimization in Eq.~\eqref{problem:EMPC} is solved again with new updated system states.

The EMPC controller in Eq.~\eqref{problem:EMPC} provides real-time optimal control action for the management of WDSs. {The implementation steps are summarized in Algorithm~\ref{alg:EMPC}. This algorithm uses mixed-integer nonlinear programming. The mass balance relationships are linear while the number of pumps is integer. The relationship between the flow and the number of pumps selected varies nonlinearly. Note that the sampling time for the EPANET simulation may be different than the one for EMPC. For example, the sampling time step for EMPC may be chosen as one hour while the sampling time step for the EPANET simulation may be chosen as five minutes.}

\begin{algorithm}[t]
\SetAlgoLined
{Set the sampling time $\Delta t$, e.g. one hour\;}
{Obtain the mathematical model of a WDS in Eq.~\eqref{eq:general model} (mass balance relationships)\;}
{Obtain the function of $\phi_j(n_j)$ in Eq. \eqref{eq:q-u function}, e.g. Eq. \eqref{eq:flow Richmond} (the approximation relating the flow to the number of pumps operating in each pump station)\;}
{Assign values to parameters: $\alpha_j$, $R$, $\underline{\mathbf{x}}$, $\overline{\mathbf{x}}$ and $\overline{\mathbf{u}}$ (electricity prices, weighting matrix in the cost function, lower and upper bounds for the water depths, maximum number of pumps available in each pump station)\;}
{Assign the total simulation time $k_{end}$ and the prediction horizon $N$\;}
 {Initialize the tank depths $\mathbf{x}(0)$\;}
 \While{{$ 0 \leq k \leq k_{end}$ (for each control time step, e.g. one hour in the EMPC)}}{
  {Obtain a sequence of water demand forecasts $\mathbf{d}(k),\ldots, \mathbf{d}(k+N - 1)$\;}
  {Solve the EMPC optimization problem in Eq.~\eqref{problem:EMPC} using mixed-integer nonlinear programming\;}
  {Select the first value of the EMPC optimal control sequence as the optimal control action as in Eq.~\eqref{eq:optimal control action}\;}
  {Reset the time step $t_l=0$ for the EPANET simulation\;}
  {Send this optimal control action to the EPANET simulator\;}
  \While{{$t_l\leq \Delta t$ (for each EPANET hydraulic time step, e.g. five minutes)}}
  {{Run the hydraulic simulation\;}
   }
   {Record the tank depths $\mathbf{x}(k+1)$ from the EPANET at the end of control time step\;}
   {Feed the tank depths $\mathbf{x}(k+1)$ back to the EMPC controller\;}
   {$k \leftarrow k+1$\;}
 }
 \caption{{EMPC of WDSs}}
 \label{alg:EMPC}
\end{algorithm}

{Furthermore, additional constraints may be required in the EMPC optimisation in Eq. (11) in order to guarantee closed-loop stability. One option is to use a constraint that forces the system onto a periodic trajectory within the control horizon if water demands are periodic as discussed by~\citeN{Wang-Salvador-2018}. This assumption of a periodic water demand does not need to hold strictly in practice. Other additional constraints such as terminal constraints, or an average performance constraint can also guarantee closed-loop stability as discussed by~\citeN{Angeli2012}. In place of additional constraints that guarantee stability, longer prediction horizons may be used \cite{Teel2005}.}

\section{Case Study: the Richmond Pruned Network}\label{section:case study}

% \subsection{Case Study Description}

\begin{figure}
	\centering
	\includegraphics[width=\hsize]{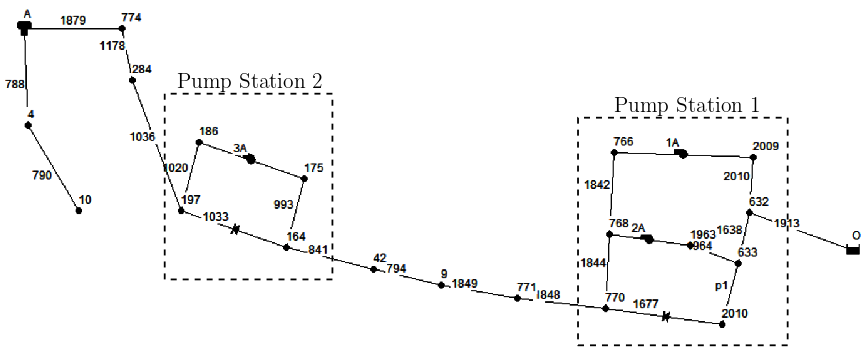}
	\caption{The layout of the Richmond Pruned network.}
	\label{fig:Richmond Pruned}
\end{figure}

\begin{figure}
	\centering
	\includegraphics[width=\hsize]{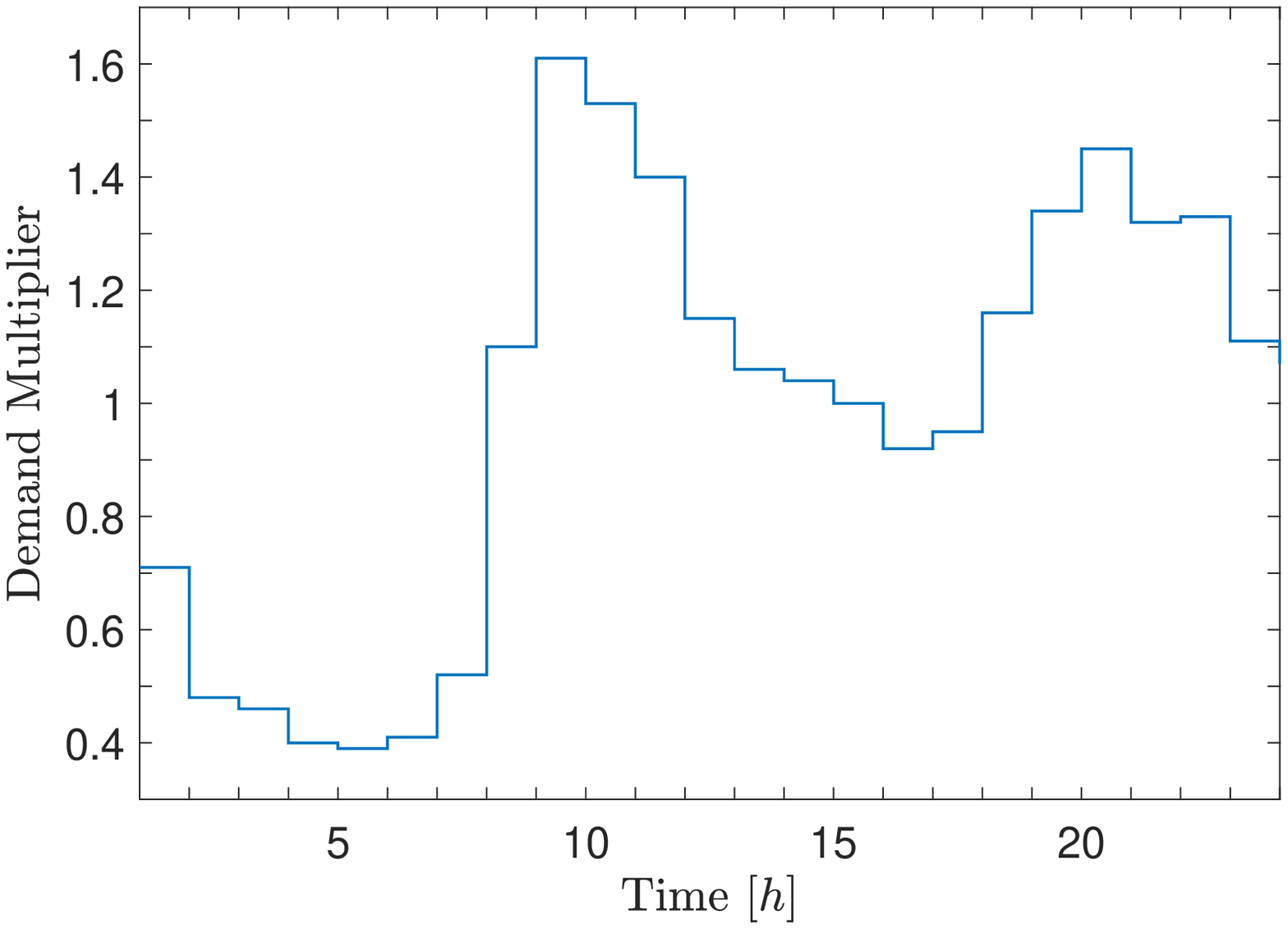}
	\caption{Demand multipliers for the Richmond Pruned network.}
	\label{fig:demand multiplier}
\end{figure}

The Richmond water network was taken from a part of the Yorkshire water supply area in the UK described in the study by~\citeN{Van2004}. Its EPANET simulation model can be found at the University of Exeter website (\url{https://emps.exeter.ac.uk/engineering/research/cws/resources/benchmarks/operation/richmond.php}). To test the proposed EMPC framework, a portion of the original Richmond network, called the Richmond Pruned network, was developed based on the Richmond skeleton model. The layout of this network is shown in Fig.~\ref{fig:Richmond Pruned}. This network consists of one water storage tank, two pump stations (one that contains two parallel pumps $ 1A $ and $ 2A $; a booster pump station that contains a single pump $ 3A $), one demand sector located at node 10. The demand multipliers $ d_{m}(k) $ within 24 hours are given in Fig.~\ref{fig:demand multiplier}, in which the average demand multiplier is $1.0$. It is assumed that the demand multipliers are repeated every 24 hours. The actual daily demand flow at node 10, $ d_{10}(k) $, is given by the product of a given base demand $ \bar{d}_{10} $ and the demand multiplier $ d_{m}(k) $, that is, $ d_{10}(k) = \bar{d}_{10} d_{m}(k) $. 

\subsection{Mathematical Model of the Richmond Pruned Network}

The mathematical model of the Richmond Pruned network is introduced in this section. Based on the EPANET results of this network, the data used to fit power and flow respectively as functions of the number of pumps operating for EMPC are reported in Table~\ref{table:fitting flow}, where the total inflow $ q_A $ to the storage tank $ A $ and power of the individual pumps for pump station 1 (PS1) and pump station 2 (PS2) are shown with different combinations of pump operations defined by the numbers of operating pumps $ n_1 $ and $ n_2 $.

\begin{table}
	\caption{{Fitting data for pump inflows to Tank A and pump power values (data from EPANET).}}
	\label{table:fitting flow}
	\centering
	\small
	\renewcommand{\arraystretch}{1.25}
	\begin{tabular}{c c | c c c c c c c c}
		\hline\hline
		 $ n_1 $ & $ n_2 $ & $ q_A $ & {PS1 Head} & {PS2 Head} & {PS1 Eff.} & {PS2 Eff.} & PS1 Power & PS2 Power & {Total Power}\\
		 & & [L/s] & {[m]} & {[m]} & & & [kW] & [kW] & {[kW]}\\
		\hline
		0 & 0 & 0 & {-} & {-} & {-} & {-} & 0 & 0 & {0}\\
		1 & 0 & 25.21 & {123.88} & {-} & {0.66} & {-} & 46.32 & 0 & {46.32}\\
		2 & 0 & 30.82  & {126.92} & {-} & {0.44} & {-} & 87.03 & 0 & {87.03}\\
		1 & 1 & 43.23  & {105.48} & {30.35} & {0.75} & {0.60} & 59.52 & 21.41 & {80.93}\\
		2 & 1 & 57.88  & {121.63} & {27.42} & {0.70} & {0.70} & 98.45 & 22.19 & {120.64}\\
		\hline\hline
		\multicolumn{10}{l}{$n_1$ - the number of pumps operating in PS1; $n_2$ - the number of pumps operating in PS2.} 
	\end{tabular}
	\normalsize
\end{table}

Based on the topology of this network as shown in Fig.~\ref{fig:Richmond Pruned}, the function in Eq.~\eqref{eq:x function} can be explicitly written as
\begin{equation}\label{eq:model-Richmond}
	x(k+1) = x(k) + \frac{\Delta t}{S_A} \left( q_A(k) - d_{10}(k) \right),
\end{equation}
where $ S_A $ is the plan area of Tank A and $ q_A $ is the inflow to Tank A. In Eq.~\eqref{eq:x function}, the units for water depth are meters ([m]) and the units for all the flow variables (inflows, outflows and water demands) are cubic meters per second ([m$^3$/s]). 

To find the total inflow $ q_A $ to Tank $ A $, a polynomial equation based on EPANET derived data can be used in Eq.~\eqref{eq:q-u function}
\begin{equation}\label{eq:flow Richmond}
	q_A(k) = c_1 n_1(k) + c_2 n_1(k)^2 + c_3 n_2(k) + c_4 n_1(k) n_2(k),
\end{equation}
where $ c_1=25.215 $, $ c_2=30.84 $, $ c_3=43.23 $ and $ c_4=57.89 $. These parameters are fitted so that Eq.~\eqref{eq:flow Richmond} accurately reproduces the flow values obtained from the EPANET as shown in Table~\ref{table:fitting flow}.

\subsection{Cost Functions and Constraints}

According to Table~\ref{table:fitting flow}, the economic cost in Eq.~\eqref{eq:general single economic cost function} can be approximated by
\begin{equation}
 	{\ell_e(\mathbf{u}(k), \alpha (k)) = \alpha (k)  (n_1(k) + n_2 (k)) p \Delta t, \qquad k=0,1,\ldots, N,}
\end{equation}
where $ \alpha(k) $ is time-varying price at time $ k $, $ \Delta t $ is the sampling time, {and $ p $ is a constant approximation of total power consumed at both pump stations based on EPANET data.} For the Richmond Pruned network, we assume an identical electricity cost price variation (i.e. the tariff of PS1, $\alpha=2.41 $ pence/kWh (UK) in the off-peak period and $\alpha=6.79 $ pence/kWh in the peak period - the values for PS1 in the original Richmond network) is used for both pump stations. Based on the EPANET results in Table~\ref{table:fitting flow}, it is obvious that the pump selection of $ n_1 =2 $ and $ n_2 = 0 $ is not a good option compared to the one of $ n_1 =1 $ and $ n_2 = 0 $ since the energy consumed is higher and the pumping flow is lower. Hence this selection is excluded in the search for optimal pump operations. {For the remaining four selections of pumps, the parameter $p$ was set as $ 40.21 $ kW based on the values in Table~\ref{table:fitting flow}.}

The weighting matrix $ R $ in Eq.~\eqref{eq:smoonthness cost function} is set as $ R = \left [\begin{smallmatrix} 100 & 0\\0 & 50
\end{smallmatrix}\right ]$. The order of the terms is chosen to provide a reasonable balance between economic cost of operation, $ \ell_e $, and the desire to reduce the turning on and off of the pumps to a level commensurate with typical operational practice. The relative weighting between the two terms is an artefact of the asymmetric network used in this case study. The weighting term for PS1 is relatively larger than the one for PS2. This is due to the fact that when Pump 3A in PS2 is turned on, one or two pumps may operate in PS1. 

Based on the original Richmond network, the water depth constraint for Tank A (i.e. the capacity of this tank) can be found in the EPANET file as
\begin{equation}
	\underline{\mathbf{x}} \leq \mathbf{x}(k) \leq  3.37,
\end{equation}
where the minimum water depth was chosen to be $ \underline{\mathbf{x}} = 1.4 $ m for the simulation {(allowing for a reserve of approximately 40\% of the full storage tank volume)}.

The control input can be set as $ \mathbf{u}(k) = \left[ n_1(k), n_2(k) \right]^{\top}  $ with the constraints
\begin{subequations}
	\begin{align}
		& \left[ 0, 0 \right] ^{\top} \leq \mathbf{u}(k) \leq \left[ 2,1 \right] ^{\top}, \label{eq:Richmond u constraint-1}\\
		& n_1(k) \geq  n_2(k),\label{eq:Richmond u constraint-2}\\
		& n_1(k) - n_2(k) \leq 1, \label{eq:Richmond u constraint-3}
	\end{align}
\end{subequations}
where Eq.~\eqref{eq:Richmond u constraint-1} indicates the available pumps in pump station 1 and 2. Eqs.~\eqref{eq:Richmond u constraint-3} ensure that the pump selection of $ n_1 =2 $ and $ n_2 = 0 $ is not used.

%%%%%%%%%%%%%%%%%%%%%%%%%%%%%%%%%%%%%%%%%%%%%%%%%%%%%%%%%%%%%%%
%%%%%%%%%%%%%%%%%%%%%%%%%%%%%%%%%%%%%%%%%%%%%%%%%%%%%%%%%%%%%%%
\section{Results}\label{section:results}

%The EMPC optimization was carried out using a laptop with an Intel i7-8550U CPU and 16GB RAM with a closed-loop platform implemented in Matlab.

The optimization problem was solved by integer programming implemented with the Yalmip toolbox~\cite{yalmip} and the Artelys Knitro solver~\cite{knitro}. The EMPC controller was connected to the EPANET hydraulic simulator. The optimal control action from the EMPC controller was sent to this simulator via the EPANET-Matlab toolkit~\cite{Eliades2016}. The EMPC controller was implemented with the sampling time $ \Delta t = 1 \; \mathrm{hour} = 3600 \; \mathrm{seconds} $ and the prediction horizon was chosen to be {$ N = 24 \;\mathrm{hours} $}. With different base demands at node 10, the closed-loop simulation results for 4 days (96 hours) are preserved and discussed in the following sections.

\subsection{Simulation Results of EMPC with EPANET}

To assess the performance of the proposed EMPC, the closed-loop simulations with the Richmond Pruned network in EPANET have been carried out from the same initial water depth $ \mathbf{x}(0) = 3.12 \mathrm{m} $ in Tank $ A $. In PS1, pumps 1A and 2A are of the same type. In simulations, we arbitrarily choose to use Pump 2A when $ n_1 = 1 $. For the demand sector at Node $ 10 $, six different water demand loading cases were chosen by setting the base demands as $ \bar{d}_{10}=5,15,25,35,45,55 $ L/s, respectively. In simulations, it was verified that with the proposed EMPC controller, this network can be operated to satisfy all the demands up to $ \bar{d}_{10} = 57.9 $ L/s but fails above this level (demands exceed capacity of the pumps and Tank A empties). For the demand loading larger than $ \bar{d}_{10} = 57.9 $ L/s, no controller or operational strategy can handle such demands since the average demand exceeds maximum pumping capacity. For the cases when $  \bar{d}_{10} > 5 $ L/s, for the step $ t \geq 8 $ over {the prediction horizon $ N $} of~\eqref{problem:EMPC}, the decision variables $\tilde{\mathbf{u}}(k),\ldots,\tilde{\mathbf{u}}(k+7)$ were treated as integer variables while the decision variables $ \tilde{\mathbf{u}}(k+8),\ldots,\tilde{\mathbf{u}}(k+N-1) $ were treated as continuous variables in order to reduce the computational burden required to find the solution.

\begin{figure}
	\centering
	\includegraphics[width=\hsize]{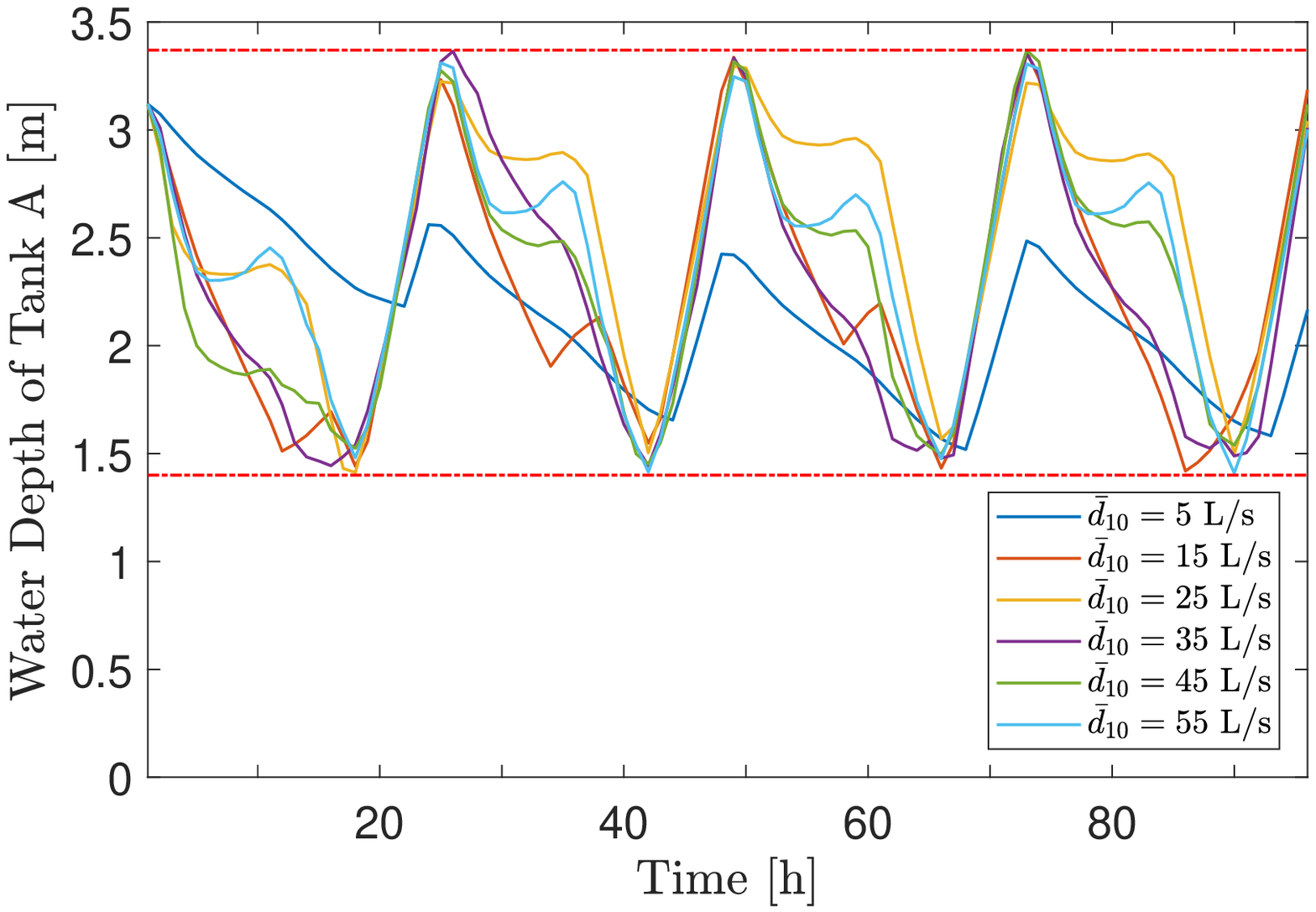}
	\caption{Water depths in Tank A for different demands at node 10.}
	\label{fig:X}
\end{figure}

For these six cases, the water depth variations of Tank A are shown in Fig.~\ref{fig:X}. From the same initial water depth, the EMPC optimization results show Tank A fills up and then decreases before arriving at the given minimum depth of $ \underline{x} = 1.4 \mathrm{m} $. In general, the tank fills up when the electricity tariff is low and empties when it is high. Since the demand multipliers follow a periodic pattern over 24 hours, the water depth variations also follow a periodic behavior for each of these six cases after the effect of the initial conditions has subsided. 

\begin{figure}
	\centering
	\subfigure[PS1 (Normalized Tariff= Tariff/4)]{\includegraphics[width=0.485\hsize]{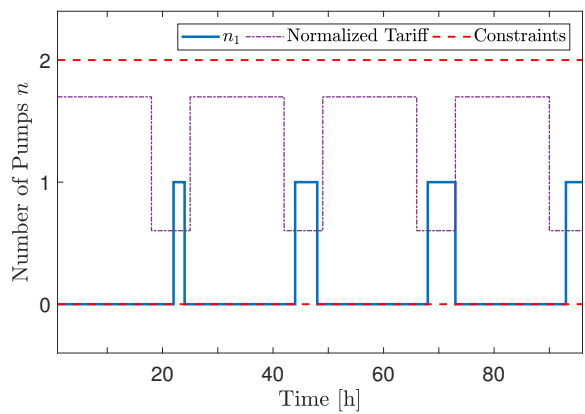}\label{fig:U-d5-a}}
	\subfigure[PS2 (Normalized Tariff= Tariff/4)]{\includegraphics[width=0.485\hsize]{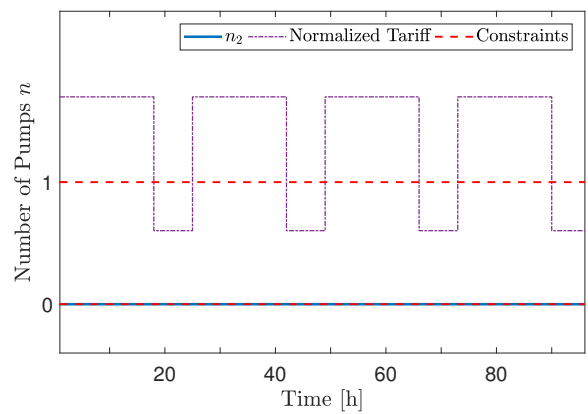}\label{fig:U-d5-b}}
	\subfigure[Water Depth in Tank A]{\includegraphics[width=0.485\hsize]{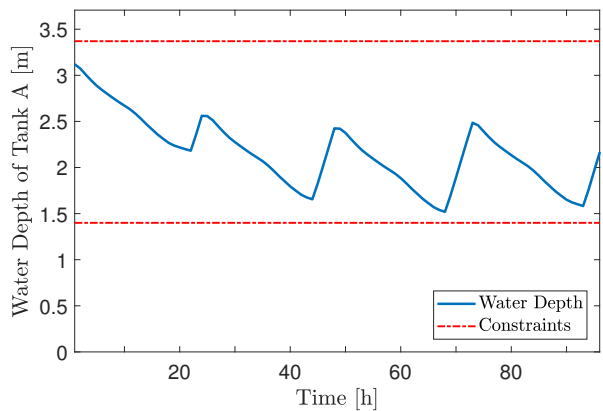}\label{fig:X-d5}}
	\caption{Optimal pump operations and variation of water depth in Tank A by the EMPC controller with $ \bar{d}_{10}  = 5 $ L/s (Only Pump 2A operates in PS1 in the off-peak tariff period).}
	\label{fig:U-d5}
\end{figure}

\begin{figure}
	\centering
	\subfigure[PS1 (Normalized Tariff= Tariff/4)]{\includegraphics[width=0.485\hsize]{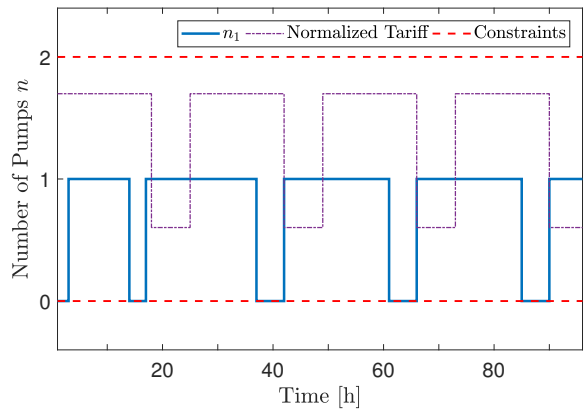}\label{fig:U-d25-a}}
	\subfigure[PS2 (Normalized Tariff= Tariff/4)]{\includegraphics[width=0.485\hsize]{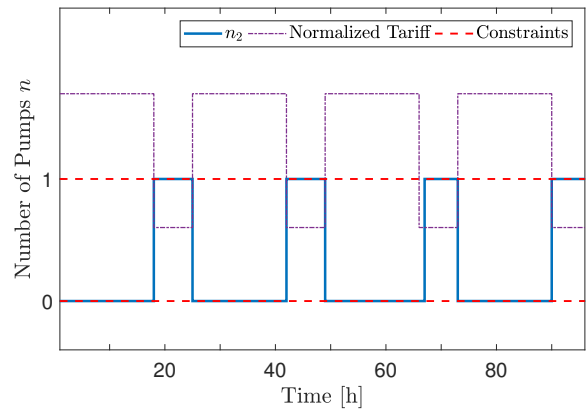}\label{fig:U-d25-b}}
	\subfigure[Water Depth in Tank A]{\includegraphics[width=0.485\hsize]{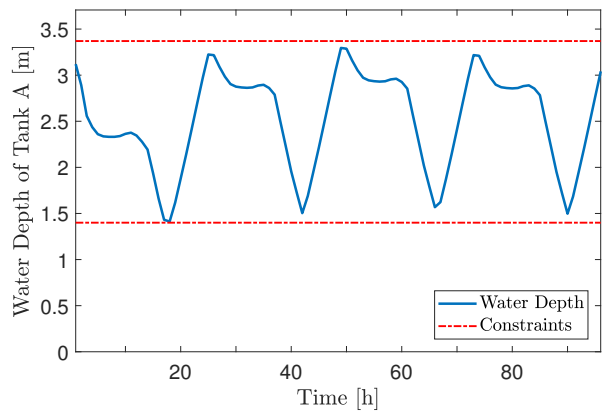}\label{fig:X-d25}}
	\caption{Optimal pump operations and variation of water depth in Tank A by the EMPC controller with $ \bar{d}_{10} =25 $ L/s (Only Pump 2A in PS1 and Pump 3A in PS2 operate. Pump 2A operates all the time in the off-peak tariff period and Pump 3A only operates in the off-peak tariff period).}
	\label{fig:U-d25}
\end{figure}

\begin{figure}
	\centering
	\subfigure[PS1 (Normalized Tariff= Tariff/4)]{\includegraphics[width=0.485\hsize]{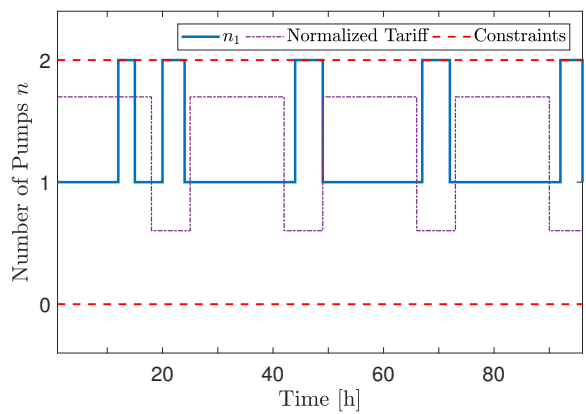}\label{fig:U-d45-a}}
	\subfigure[PS2 (Normalized Tariff= Tariff/4)]{\includegraphics[width=0.485\hsize]{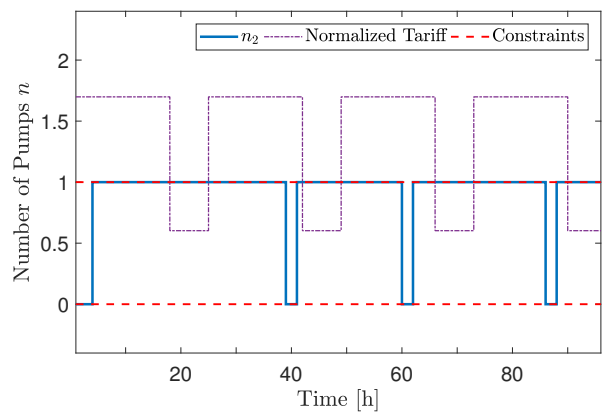}\label{fig:U-d45-b}}
	\subfigure[Water Depth in Tank A]{\includegraphics[width=0.485\hsize]{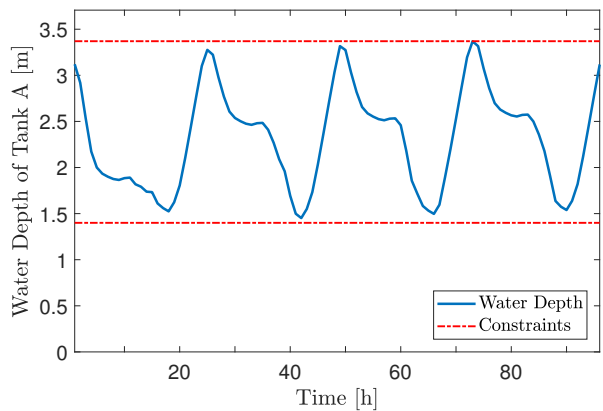}\label{fig:X-d45}}
	\caption{Optimal pump operations and variation of water depth in Tank A by the EMPC controller with $ \bar{d}_{10} = 45 $ L/s (All 3 pumps operate at various times).}
	\label{fig:U-d45}
\end{figure}

For $ \bar{d}_{10}  = 5, 25,45 $ L/s, the resulting pump operations are shown in Fig.~\ref{fig:U-d5} to Fig.~\ref{fig:U-d45}. As shown in Fig.~\ref{fig:U-d5}, since the base demand is small, only one pump in PS1 is required, and pumping only takes place in the off-peak tariff period as shown in Fig.~\ref{fig:U-d5-a}. When the demand is increased to $ \bar{d}_{10}  = 25 $ L/s, operation of one pump in PS1 is not enough to provide the required flows and pumping in PS2 is required, but only in the off-peak tariff period as shown in Fig.~\ref{fig:U-d25-b} while some pumping in PS1 is necessary in the peak tariff period. However, only one pump (out of 2 pumps) in PS1 is operated and this pump is switched off for some of the time during the peak tariff period. When the demand is increased to $ \bar{d}_{10}  = 45 $ L/s, both pump stations are required to be operated in order to provide enough water. From the optimal solution of EMPC, a single pump in PS1 is operated during the peak tariff period and two pumps in PS1 are operated in the off-peak tariff period as shown in Fig.~\ref{fig:U-d45-a}. For PS2, the single pump is needed for most of the time and is only switched off for some of the time during the peak tariff period. 

In the cases in Figs.~\ref{fig:U-d5} and~\ref{fig:U-d25}, most of the pumping takes place in the off-peak tariff period and no unexpected pump switches happened. For the case when $ \bar{d}_{10}  = 45 $ L/s, the optimal pump operations are shown in Figs.~\ref{fig:U-d45-a}-\ref{fig:U-d45-b} and the water depth variation of Tank A is shown in Fig.~\ref{fig:X-d45}. From the beginning, Tank A drains since the demand is high and Pump 1A in PS1 is not turned on since it is in the peak tariff period. As the water depth in Tank A gets close to the lower limit, both Pumps 1A and 2A in PS1 are eventually turned on in order to avoid the water depth going below 1.4m but Pump 1A is only used for a short period of time in the peak tariff period but used most of time in the off-peak tariff period. Pump 3A in PS2 still operate until around $ k=40 $. Since it is still in the peak tariff period and there is enough water in Tank A, Pump 3A is turned off for some time and turned on again in the off-peak period.

To further assess the performance of the proposed EMPC, the total pumped water volume, the total consumed energy, and the total economic cost are calculated for six water demand loading cases and the computational results are reported in Table~\ref{table:comparison result-EMPC}. Note that the results of the economic costs are actual pumping energy costs by implementing the chosen pump operations. As the volume of pumped water increases, the consumed energy also increases as well as the total cost. The total cost per unit volume is also computed in Table~\ref{table:comparison result-EMPC}. Since more water is pumped in the peak tariff period, the total cost per unit volume also increases as the water demand increases. Furthermore, the average pump efficiencies for pumps $ 1A $, $ 2A $ and $ 3A $ were obtained from the EPANET simulator and reported in Table~\ref{table:comparison result-EMPC}. With the proposed EMPC controller, all the average pump efficiencies are around 70\%.

% \begin{table}
% 	\caption{Computation results with the EMPC.}
% 	\label{table:comparison result-EMPC}
% 	\centering
% 	\small
% 	\renewcommand{\arraystretch}{1.25}
% 	\begin{tabular}{c | c c c c c c c c c c}
% 		\hline\hline
% 		$ \bar{d}_{10} $ & Tot. Vol. Pumped & Tot. Energy & Tot. Cost  & Cost per m$^3 $ & $ \bar{\eta}_{1A} $  & $  \bar{\eta}_{2A} $  & $ \bar{\eta}_{3A} $\\
		
% 		[L/s] & [m$^3$] & [kWh] & [\pounds] & [\pounds/m$^3$] & [\%]  & [\%] & [\%]\\
% 		\hline
% 		5  & {1398} & 712 & 17.16  & 0.0134 & 0 & 66.67 & 0\\
% 		15 & {5276} & 2712 & 96.69 & 0.0200 & 0 & 70.17 & 60.24\\
% 		25 & {8693} & 4493 & 207.28  & 0.0260 & 0 & 68.74 & 60.25\\
% 		35 & {12161} & 6427 & 319.24  & 0.0286 & 69.71 & 68.83 & 64.04\\
% 		45 & {15580} & 8327 & 444.01  & 0.0311 & 69.79 & 72.97 & 62.83\\
% 		55 & {19009} & 10795 & 595.67 & 0.0342 & 69.75  & 70.88 & 68.37 \\
% 		%\hline
% 		%\multicolumn{2}{l}{$\ast$ $D_{50}$ represents the median particle diameter} \\
% 		\hline\hline
% 	\end{tabular}
% 	\normalsize
% \end{table}

\begin{table}
	\caption{Computation results with EMPC for 4-day simulation.}
	\label{table:comparison result-EMPC}
	\centering
	\small
	\renewcommand{\arraystretch}{1.25}
	\begin{tabular}{c | c c c c c c c c c c}
		\hline\hline
		$ \bar{d}_{10} $ & {Tot. Vol.$^\ast$} & Tot. Energy & Tot. Cost  & Cost per m$^3 $ & $ \bar{\eta}_{1A} $  & $  \bar{\eta}_{2A} $  & $ \bar{\eta}_{3A} $\\
		
		[L/s] & [m$^3$] & [kWh] & [\pounds] & [\pounds/m$^3$] &  &  & \\
		\hline
		5  & {1398} & 712 & 17.16  & 0.0134 & - & 0.67 & -\\
		15 & {5276} & 2712 & 96.69 & 0.0200 & - & 0.70 & 0.60\\
		25 & {8693} & 4493 & 207.28  & 0.0260 & 0 & 0.69 & 0.60 \\
		35 & {12161} & 6427 & 319.24  & 0.0286 & 0.70 & 0.69 & 0.64\\
		45 & {15580} & 8327 & 444.01  & 0.0311 & 0.70 & 0.73 & 0.63\\
		55 & {19009} & 10795 & 595.67 & 0.0342 & 0.70  & 0.71 & 0.68\\
		%\hline
		\hline\hline
		\multicolumn{8}{l}{{$^\ast$Tot. Vol. is the total water volume delivered to Tank A.}} \\
	\end{tabular}
	\normalsize
\end{table}

%\subsection{Comparison with Open-loop Implementation}
%
%The simulations with the EMPC controller and the EPANET hydraulic simulator were carried out with $ \bar{d}_{10} = 5 $ L/s both using open-loop and closed-loop strategies as discussed before. The open-loop and closed-loop water depths are compared as shown in Fig.~\ref{fig:Comparison OL-CL}. Both cases are from the same initial water depth $ \mathbf{x}(0) = 3.12 \mathrm{m} $. The difference is that in the open-loop implementation, there is not any feedback from the EPANET simulator for the EMPC optimization problem reinitialization. In contrast, for the closed-loop implementation, the current water depth (state state) at each time $ k $ (based on EPANET results) is fed back to the EMPC controller as the first predicted state. In Fig.~\ref{fig:Comparison OL-CL}, the closed-loop result shows that water depth is regulated in a potential periodic manner, however, in the open-loop implementation more water is transferred into Tank A. Consequently, Tank A will be full leading to the infeasibility of the EMPC optimization problem. So the closed-loop implementation is the appropriate way for implementing the proposed EMPC framework.

%\begin{figure}
%	\centering
%	\includegraphics[width=\hsize]{Figs/Comparison_OL_CL.eps}
%	\caption{Comparison of the open-loop and closed-loop implementation.}
%	\label{fig:Comparison OL-CL}
%\end{figure}

\subsection{Comparison with Trigger-Level Control}

\begin{figure}
	\centering
	\subfigure[Total Pumping Flow]{\includegraphics[width=0.485\hsize]{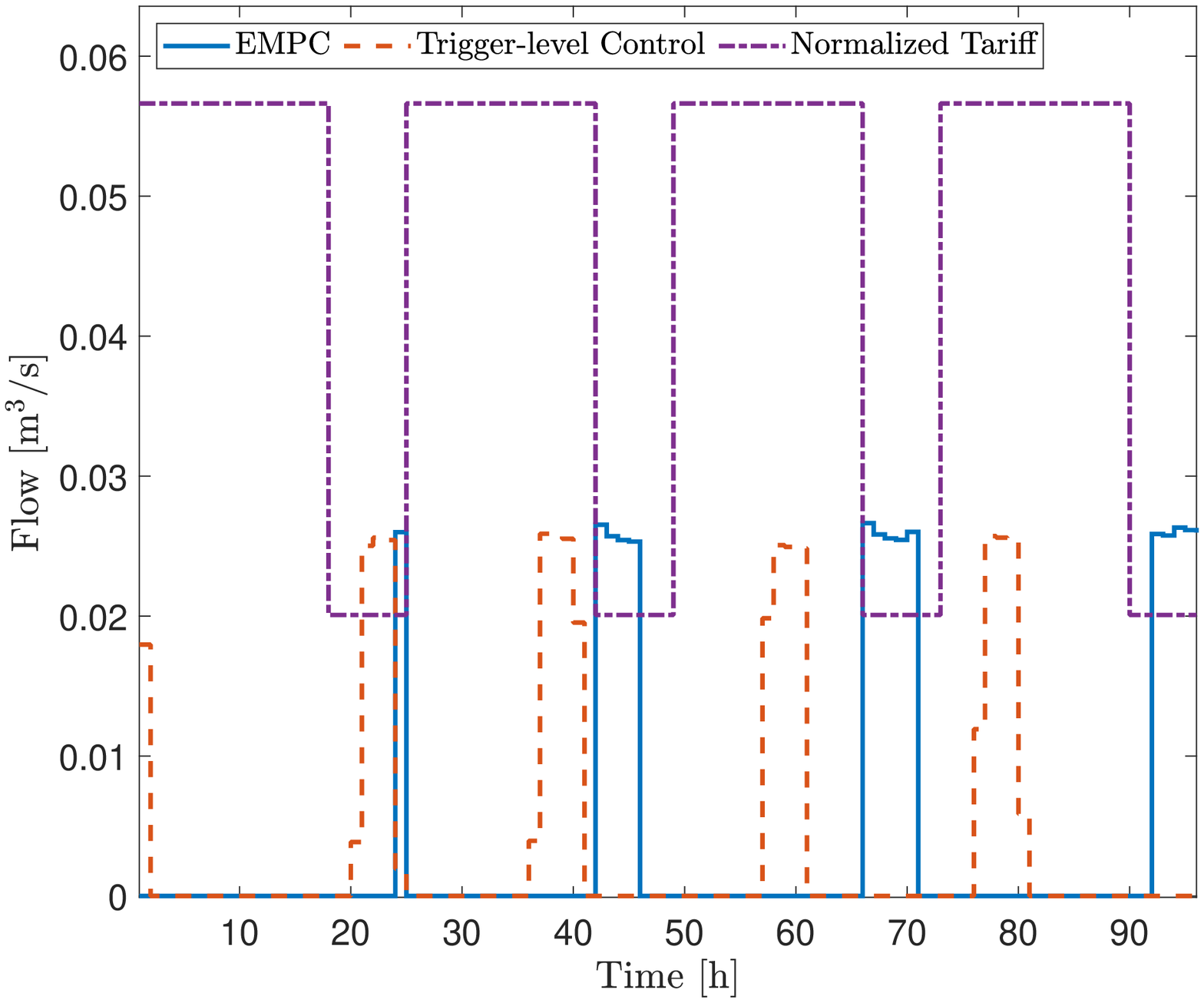}\label{fig:Comparison-a}}
	\subfigure[{Cumulative Water Volume}]{\includegraphics[width=0.485\hsize]{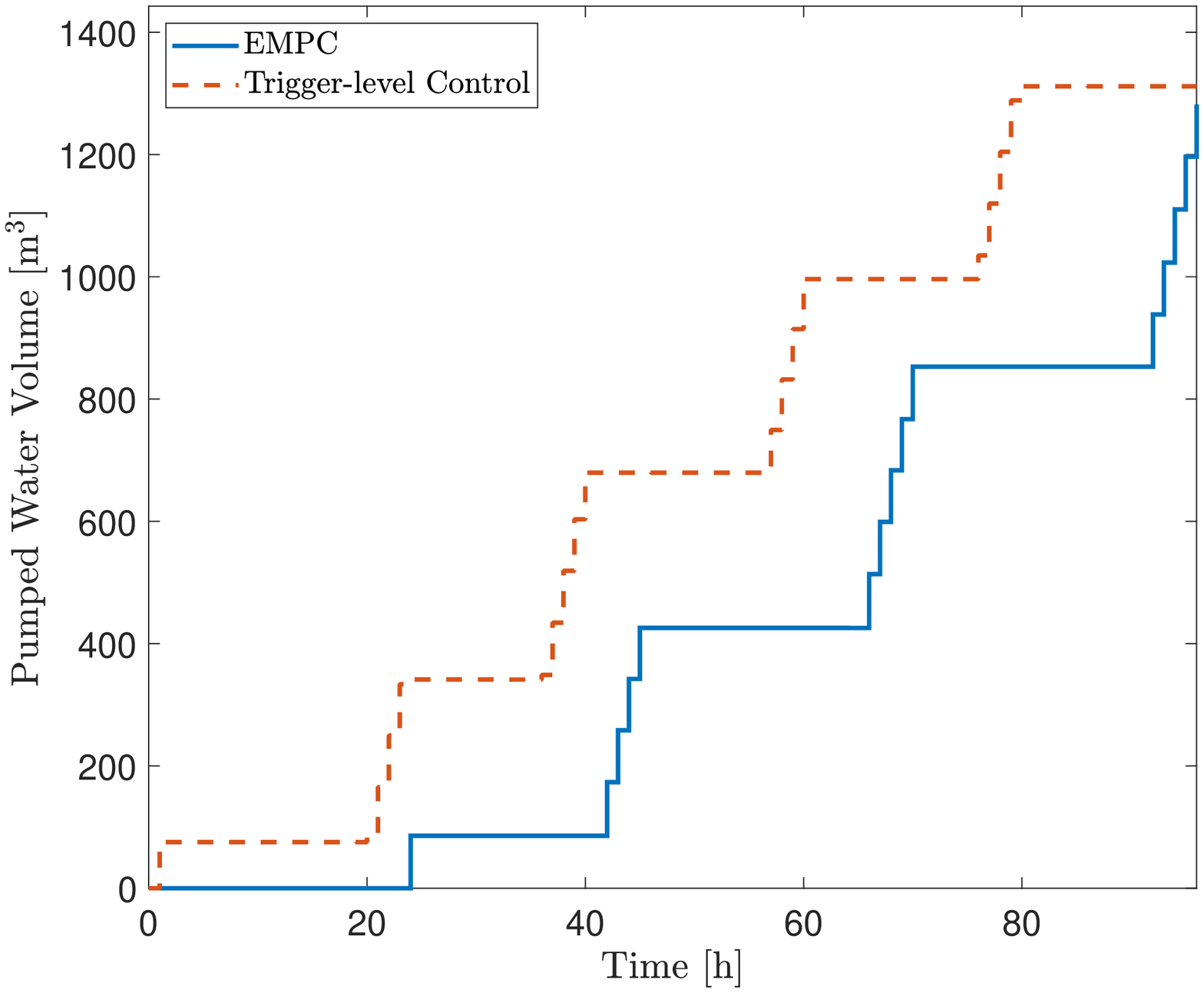}\label{fig:Comparison-b}}\\
	\subfigure[{Cumulative Energy}]{\includegraphics[width=0.485\hsize]{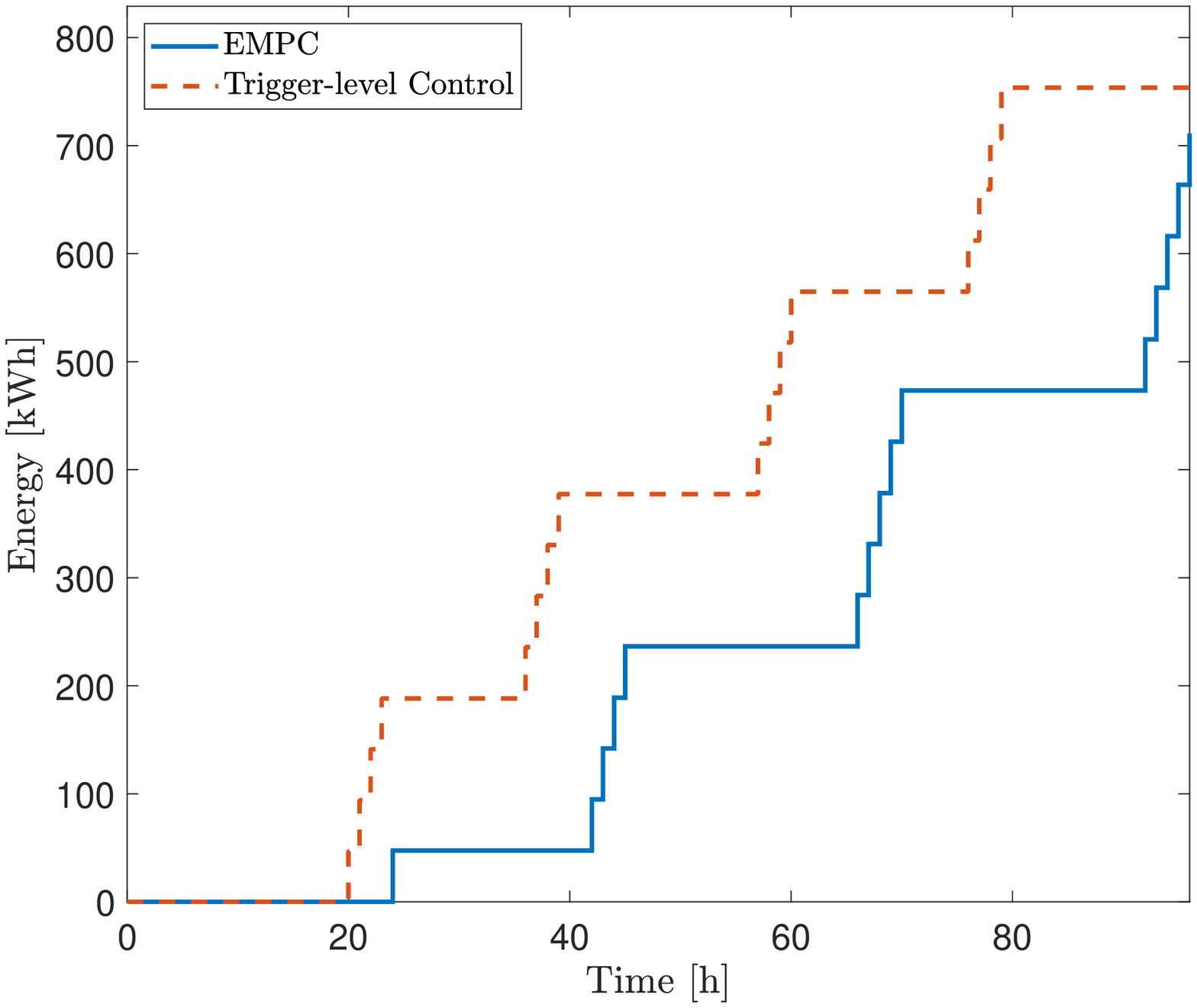}\label{fig:Comparison-c}}
	\subfigure[{Cumulative Cost}]{\includegraphics[width=0.485\hsize]{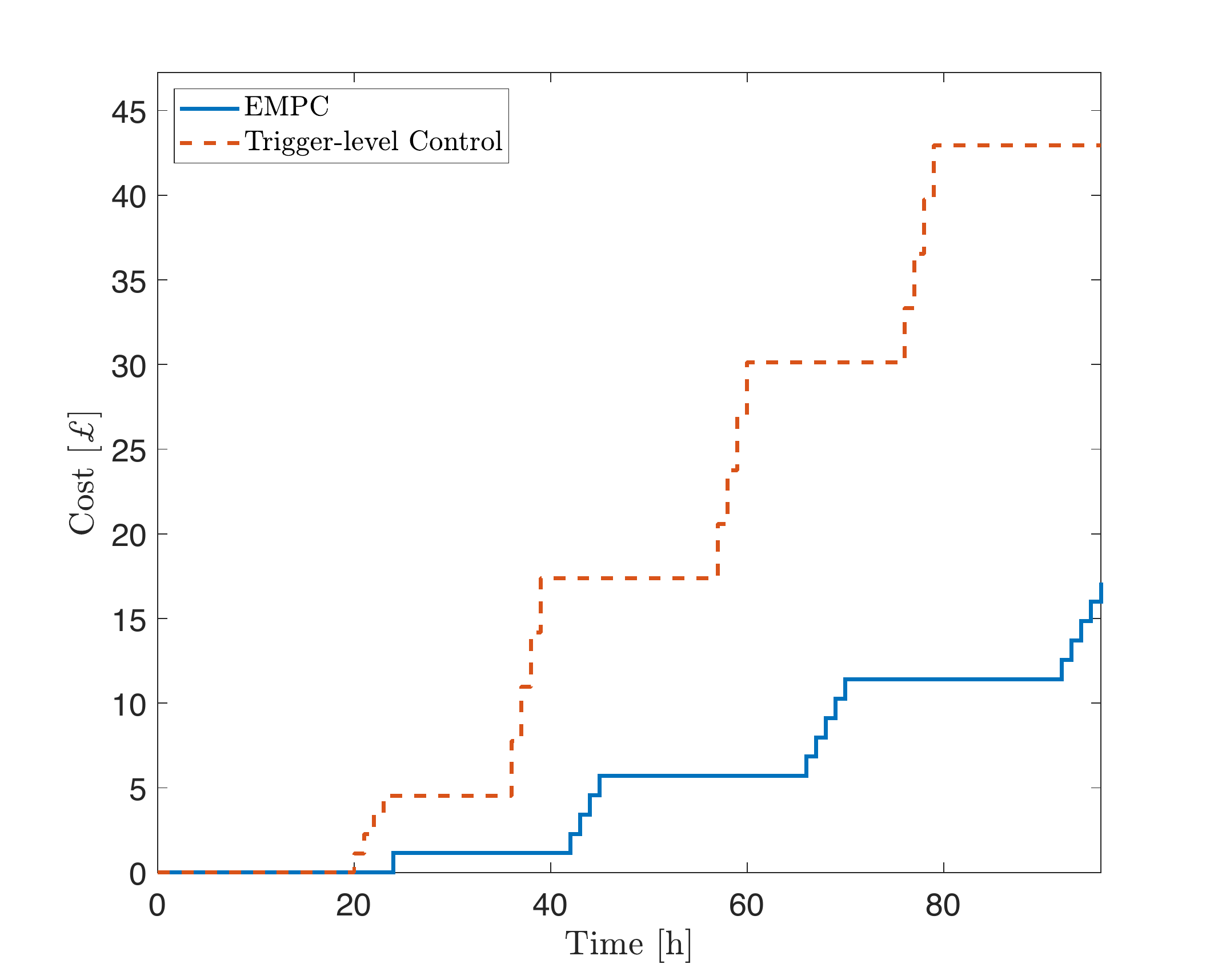}\label{fig:Comparison-d}}\\
	\caption{Comparison of EMPC and trigger-level control with $ \bar{d}_{10}  = 5 $ L/s (Note for EMPC, Pump 2A operates only in the off-peak tariff period).}
	\label{fig:Comparison}
\end{figure}

To compare the performance of EMPC with a traditional control strategy, the traditional trigger-level control is also applied for the Richmond Pruned network. The trigger levels for the three pumps are shown in Table~\ref{table:trigger level controller}. With $ \bar{d}_{10} = 5 $ L/s, the results are compared in Fig.~\ref{fig:Comparison}. At the beginning, the initial tank water depth is $ \mathbf{x}(0) = 3.12 \mathrm{m} $. Based on Table~\ref{table:trigger level controller}, Pump 2A was turned on. Since trigger-level control does not take into account the time-varying electricity price, the pumping flows will occur at any time based on the water depth in Tank A without the regard for the peak and off-peak tariff periods as shown in Fig.~\ref{fig:Comparison-a}. In contrast, for EMPC, the pumps are only operated in the off-peak tariff period for $ \bar{d}_{10} = 5 $ L/s. Consequently, as shown in Fig.~\ref{fig:Comparison-b} and~\ref{fig:Comparison-c}, the total pumped water volume and the total energy consumed are similar for both EMPC and trigger-level control while the cost with EMPC is significantly lower than the cost with trigger-level control.

\begin{table}
	\caption{The trigger-level controller setup.}
	\label{table:trigger level controller}
	\centering
	\small
	\renewcommand{\arraystretch}{1.25}
	\begin{tabular}{c | c c}
		\hline\hline
		Pump & Trigger Level$^{*}  $ - ON  & Trigger Level - OFF \\
		 & [m] & [m]\\
		\hline
		$ 1A $ & $ \leq 2.37 $ & $ \geq 2.98 $\\
		$ 2A $ & $ \leq 1.40 $ & $ \geq 3.25 $ \\
		$ 3A $ & $ \leq 1.90 $ & $ \geq 3.11 $ \\
		\hline\hline
		\multicolumn{3}{l}{$^{*} $ Tank A; Lower trigger level values for {Pumps $2A$ and $3A$} were}\\ 
		\multicolumn{3}{l}{reduced compared to the original Richmond EPANET input file.} 
	\end{tabular}
	\normalsize
\end{table}

% Table discussions
The results obtained from trigger-level control with six different demand loading cases are reported in Table~\ref{table:comparison result-trigger level}. The cost ratio between EMPC and trigger-level control reported in Table \ref{table:comparison result-trigger level} is computed by
\begin{equation}
\text{Cost Ratio} = \frac{\text{Trigger-level Cost per m}^3}{\text{EMPC Cost per m}^3}.
\end{equation}

According to the cost ratio results in Table~\ref{table:comparison result-trigger level}, the optimal pump operations with EMPC has gained an economic benefit between 3\% and 250\% compared to trigger-level control. The relative benefits are more significant at lower demand flows, as there are more opportunities for intelligent scheduling to increase economic efficiency. On the other hand, at the higher demands, since there is less opportunity for improving the economic performance, the utilization of the pumps is very high and they must be operated during the high tariff period in order to satisfy demands. 

\begin{table}
	\caption{Computation results with trigger-level control for
	4-day simulation.}
	\label{table:comparison result-trigger level}
	\centering
	\small
	\renewcommand{\arraystretch}{1.25}
	\begin{tabular}{c | c c c c c}
		\hline\hline
		$ \bar{d}_{10}$ & {Tot. Vol.$^\ast$} & Tot. Energy & Tot. Cost & Cost per m$^3 $ & Cost Ratio\\
		
		[L/s] & [m$^3$] & [kWh] & [\pounds] & [\pounds/m$^3$] & \\
		\hline
		5  & {1430}  & 753 & 42.96 & 0.0328 & 2.50\\
		15 & {5088} & 2540 & 150.17 & 0.0322 & 1.55 \\
		25 & {8534} & 4385 & 242.34  & 0.0310 & 1.16 \\
		35 & {11849} & 6477 & 410.32 & 0.0378 & 1.28 \\
		45 & {15761} & 8746 & 515.52 & 0.0357 & 1.16 \\
		55 & {19093} & 10899 & 618.74 & 0.0354 & 1.03 \\
		\hline\hline
		\multicolumn{6}{l}{{$^\ast$Tot. Vol. is the total water volume delivered to Tank A.}} \\
		\multicolumn{6}{l}{Cost Ratio = Trigger-level Cost per m$^3$/EMPC Cost per m$^3$.}\\
	\end{tabular}
	\normalsize
\end{table}

{In Table~\ref{table:comparison result Day 4}, another comparison has been made for results with EMPC and trigger-level control on Day 4. For the six different demand loading cases, the cost and the cost per m$^3$ on Day 4 with trigger-level control are more expensive than with EMPC. In particular, for the cases of $\bar{d}_{10} = 5 $ and 55 L/s, more water is delivered to Tank A with EMPC but the cost with EMPC is cheaper than with trigger-level control.}

\begin{table}
	\caption{{Comparison of Day 4 (D4) for EMPC and trigger-level control.}}
	\label{table:comparison result Day 4}
	\centering
	\small
	\renewcommand{\arraystretch}{1.25}
	\begin{tabular}{c | c c c c || c c c c}
		\hline\hline
		& \multicolumn{4}{c||}{{EMPC}} & \multicolumn{4}{c}{{Trigger-level Control}}\\
		\cline{2-9}
		$ \bar{d}_{10}$ & {D4 Vol.$^\ast$} & {D4 Energy} & {D4 Cost} & {D4 Cost per m$^3$} & {D4 Vol.$^\ast$} & {D4 Energy} & {D4 Cost} & {D4 Cost per m$^3$}\\
		
		[L/s] & [m$^3$] & [kWh] & [\pounds] & [\pounds/m$^3$] & [m$^3$] & [kWh] & [\pounds] & [\pounds/m$^3$]\\
		\hline
		5  & {429} & {238} & {5.73} &  {0.0134} & {346} & {189} & {12.84} & {0.0371}\\
		15 & {1209} & {677} & {22.62} & {0.0187} & {1313} & {659} & {38.64} & {0.0294}\\
		25 & {1923} & {1084} & {48.78} & {0.0254} & {2216} & {1134} & {62.60} & {0.0282}\\
		35 & {2765} & {1560} & {82.20} & {0.0297} & {3024} & {1573} & {102.33} & {0.0338}\\
		45 & {3560} & {2086} & {108.54} & {0.0305} & {4043} & {2186} & {128.02} & {0.0317}\\
		55 & {4359} & {2710} & {151.47} & {0.0347} & {4285} & {2683} & {154.40} & {0.0360}\\
		\hline\hline
		\multicolumn{8}{l}{{$^\ast$D4 Vol. is the total water volume delivered to Tank A on Day 4.}}
	\end{tabular}
	\normalsize
\end{table}

%%%%%%%%%%%%%%%%%%%%%%%%%%%%%%%%%%%%%%%%%%%%%%%%%%%%%%%%%%%%%%%
%%%%%%%%%%%%%%%%%%%%%%%%%%%%%%%%%%%%%%%%%%%%%%%%%%%%%%%%%%%%%%%
\section{Conclusions}\label{section:conclusions}

In this study, a novel EMPC framework has been proposed for real-time operational management of WDSs. The optimal pump operations are chosen by minimizing pumping energy costs approximated with a flow-based model of WDSs. {To demonstrate the utility and advantage of this proposed EMPC framework, the Richmond Pruned network case study including three pumps in two pump stations has been used.} The closed-loop simulation with an EMPC controller and an EPANET hydraulic simulator has shown that the proposed EMPC framework is effective and efficient in finding a set of optimal pump operations with minimum pumping energy costs taking into account time-varying electricity prices. Less pumping occurs during the peak tariff period and only when it is necessary while more pumps are operated in the off-peak period. The performance of EMPC has also been compared to the tradition operational control based on trigger level values. From this comparison, under the smallest demand loading case, the energy consumption obtained with EMPC and the trigger-level control is similar at the end of the simulation time but the pumping energy is 2.5 times more expensive with trigger-level control than with EMPC. The significantly lower cost obtained with EMPC compared to the trigger-level control is due to the fact that EMPC takes into account the time-varying electricity prices. 

Uncertainties {exist} in both the mathematical model of the WDS and the forecasts of demands and electricity prices. The EMPC strategy has displayed some inherent robustness against uncertainty since it uses a simplified model when carrying out the optimization while the hydraulic simulation of a WDS was carried out using the EPANET simulator. However, robust control strategy for WDSs with enhanced robustness against uncertainty is a topic for future research.

{As a final point, as the complexity of the network increases, the computational tractability of the proposed algorithm may become challenging. In such circumstances, it may be necessary to consider distributed implementations of the algorithms to show the computational burden among the available computational resources.}

%%%%%%%%%%%%%%%%%%%%%%%%%%%%%%%%%%%%%%%%%%%%%%%%%%%%%%%%%%%%%
%%%%%%%%%%%%%%%%%%%%%%%%%%%%%%%%%%%%%%%%%%%%%%%%%%%%%%%%%%%%%
\section*{Data Availability}

Some data, models, or code generated or used during the study are available in the following online repository in accordance with funder data retention policies \url{https://github.com/yewangunimelb/watersystem}. 

\bibliography{RichmondPruned}

\end{document}